\documentclass[aps,preprint,showpacs]{revtex4}
\usepackage{graphics,rotate}

\begin{document}

\title{Constant-time solution to the Global Optimization Problem\\
using Br\"uschweiler's ensemble search algorithm}
\author{V. Protopopescu}
\email{protopopesva@ornl.gov}
\author{C. D'Helon}
\author{J. Barhen}
\affiliation{Center for Engineering Science Advanced Research, Computer Science and Mathematics Division,\\ Oak Ridge National Laboratory, Oak Ridge, TN 37831-6355 USA}

\begin{abstract}
A constant-time solution of the continuous Global Optimization Problem (GOP) is obtained by using an ensemble algorithm. We show that under certain assumptions, the solution can be \textit{guaranteed} by
% which forbid an arbitrarily high complexity, the continuous Global Optimization Problem (GOP) can be solved efficiently for any objective function, independent of dimensionality. For any GOP that satisfies the assumptions, the number of steps in computing the solution grows logarithmically with the size of the function's domain, i.e., linearly as a function of dimensionality. The assumptions allow the 
mapping the GOP onto a discrete unsorted search problem, whereupon Br\"uschweiler's ensemble search algorithm 
%[Phys. Rev. Lett. \textbf{85} 4815 (2000)] 
is applied. For adequate sensitivities of the measurement technique, the query complexity of the ensemble search algorithm depends linearly on the size of the function's domain.
%the scaling of the signal-to-noise ratio with the number of experimental trials. 
Advantages and limitations of an eventual NMR implementation are discussed.
\end{abstract}

\maketitle
\newpage

\section{The Global Optimization Problem}
\label{Introduction}

Optimization problems are ubiquitous and extremely consequential.  Their theoretical and practical interest has continued to grow from the first recorded instance of Queen Dido's problem \cite{S} to present day forays into complexity theory and large scale logistics applications (see Refs. \cite{TZ,HT,HHP,FP,DS} and references therein).

The formulation of the Global Optimization Problem (GOP) is deceptively simple: find the absolute minimum of a given function - called the objective function - over the allowed range of its variables. Sometimes, the function whose global minimum is to be found is not specified in analytic form and must be evaluated by a computer program. When we can only access the output of the computation, the program acts as a black-box tool, called an {\it oracle}.

Of course, the brute force approach of evaluating the function on its whole domain is either impossible if the variables are continuous, or prohibitively expensive if the variables are discrete, but take values in large domains and in high dimensional spaces. Since, in general, each oracle invocation (function evaluation) involves an expensive computational process, the number of function evaluations needs to be kept to a minimum.  It is not surprising that, together with accuracy, this number probably provides the paramount criterion in comparing the efficiency of competing optimization algorithms.

The primary difficulty in solving GOPs stems from the fact that the condition for determining minima, namely annulment of the gradient of the objective function, is only \textit{necessary} (the function may have another type of critical point) and \textit{local} i.e. it does not distinguish between local and global minima. The generic strategy to find the global minimum involves two main operations, namely: (i) descent to a local minimum and (ii) search for a new descent region. However, this generic strategy is powerless for certain problems. The following one-dimensional example (the golf course problem) illustrates the difficulties. Define the function $f :[0,1]$ by
\begin{eqnarray}
f(x) =  \left\{ \begin{array}{l}
~~0 ~\mbox{for}~ 0 \leq x \leq a - \epsilon/2 ~\mbox{and}~ a + \epsilon/2 \leq x \leq 1 \\
-1 ~\mbox{for}~ a - \epsilon/2 \leq x \leq a + \epsilon/2\\
%0 ~\mbox{for}~ a + \epsilon/2 \leq x \leq 1
\end{array} \right.
\end{eqnarray}
\noindent
where $0<\epsilon\ll 1$, and $a$ is a point in the interval $(\epsilon/2, 1-\epsilon/2)$, but is otherwise unknown.  From this definition, it is clear that to find the absolute minimum of $f$, one should evaluate it within the $\epsilon$ well-interval around the unknown point $a$.  If this function is defined like an oracle, i.e. if one does {\it not} know its definition above and the value of the number $a$, the probability $P_\epsilon$ of choosing the variable $x$ within the well-like interval where $f(x)=-1$, is $\epsilon$. In the $n$-dimensional version of this problem, the probability $P_\epsilon$ becomes $\epsilon^n$, so the complexity of the problem grows exponentially with $n$ (the dimensionality curse).  Moreover, knowledge about the (partial) derivative(s) of $f$ would not help, since they are all zero whenever defined.
In the light of this example, it seems that without additional information about the structure of the function there is no reasonable expectation of deciding upon an efficient optimization strategy and one is left with either strategies that have limited applicability or the exhaustive search option.

Thus, new approaches are needed, which take advantage of \textit{additional information to reduce the complexity of the problem} to a manageable level. We emphasize that this information is actually available in some classes of GOP - for instance it happens to be available for the golf course problem above - but \textit{cannot} be taken advantage of within conventional optimization algorithms. 

%Using this additional information enables one to limit the complexity of the objective function, so that the ``hardness'' of the GOP becomes independent of the actual objective function.

%Here we present an approach that uses this information efficiently to map the continuous GOP into a discrete search problem (DSP). Once this mapping is completed, we apply a variant of Br\"uschweiler's search algorithm, as described in the following section, to obtain a value within the basin of attraction of the global minimum in a number of function evaluations that scales logarithmically with respect to the cardinality of the resulting discrete set. The parallelism of ensemble computing can provide an ``exponential speed-up'', however this demands an exponentially large number of processors. Thus, we have used the more modest expression ``constant-time'', to express that the search problem can be solved in a constant time, given exponential resources. We can only claim that the approach in this Letter may be faster than other algorithms, for problems up to some limited size.

Here we present an approach that uses this information to map the continuous GOP into a discrete search problem (DSP). Once this mapping is completed, we apply a variant of Br\"uschweiler's search algorithm, as described in the following section, to obtain a value within the basin of attraction of the global minimum in a number of function evaluations that scales logarithmically with respect to the cardinality of the resulting discrete set. While the parallelism of ensemble computing indeed provides an ``exponential speed-up'', it also demands an exponentially large number of processors. Thus, we use the phrase ``constant-time'', to account for the relationship between speed and resources.

\section{Mapping the continuous GOP to a discrete search problem}
\label{Mapping}

Consider a real function of $d$ variables, $f(\textbf{x})$, $\textbf{x}=(x_{1},x_{2},\dots,x_{d})$ that has a certain degree of smoothness. The precise definition is not critical for the argument below. Without restricting generality, we can assume that $f$ is defined on $[0,1]^{d}\rightarrow [0,1]$.  Assume now that: (i) there is a unique global minimum and its value is zero; (ii) there are no local minima whose value is infinitesimally close to zero; in other words, the values of the other minima are larger than a constant $\delta > 0$, and (iii) the size of the basin of attraction for the global minimum measured at height $\delta$ is known to be $\textbf{r}_\delta$. We note that these assumptions can be actually verified for the problem at hand, if additional information is available. Also, these assumptions may in general be relaxed to some extent, as outlined in the first part of the Discussion.

The implementation idea is the following. Instead of $f(\textbf{x})$, consider the function $g(\textbf{x}) := (f(\textbf{x}))^{1/m}$.  For sufficiently large $m$, this function will take values very close to one, except in the vicinity of the global minimum, which will maintain its original value, namely zero.  Of course, other transformations may be used to achieve essentially the same result.  Now divide the hypercube $[0,1]^d$ in small $d$-dimensional hypercubes with sides $\Delta x = 1/M$, where $M$ is a positive integer.  At the midpoint of each of these hypercubes, define the function $h(\textbf{x}):=1-INT[g(\textbf{x}) + 1/2]$ ($INT$ denotes the integer part).
%The function $h(\textbf{x})$ is defined on a discrete set and takes only values one and zero.

%Without loss of generality, we can take $\epsilon = 1/N$ and divide the segment $[0,1]$ into $N$ equal intervals.  By evaluating the function at the midpoint of the $N$ intervals, we obtain a discrete function that is equal to one in $N-1$ points and equal to zero in one point, which is equivalent to a DSP.

As a result of the transformations above, we reduced the GOP to a DSP for $h:\{1,2,\dots,N\}\rightarrow\{0,1\}$, which is an integer-valued function defined on a discrete set of $N:=2^{n+1}$ points. The function $h(i)$ is known to be zero for all inputs $i$, except for one special input $i=q$, where $h(q)=1$. Here $N = M^d$ and the domain on which the function $h$ is zero depends on $m$ and on the size, $\textbf{r}_\delta$, of the basin of attraction of the global minimum at height $\delta$. Thus the problem becomes to find the value of the special input $q$ efficiently. Recently, we proposed an adaptation of Grover's quantum search algorithm to solve the continuous GOP \cite{Protopopescu02}, which requires $O(\sqrt N)$ function evaluations \cite{grover,Nayak98,Beals98}.

In this paper, we propose to further improve the efficiency, by using an ensemble search algorithm, which is described in detail in the following section. The algorithm is adapted from Br\"uschweiler's ensemble search algorithm \cite{Madi98,Bruschweiler00}, using a parallel technique to achieve exponential speedups for problems up to a certain size, dictated by the sensitivity of the measurement process. As mentioned before, efficiency is understood in relation to the query complexity of the algorithm. Indeed, when $N$ is large and the function evaluation (i.e. the oracle) is costly in terms of computational complexity, reducing the number of function evaluations is critical.

Application of the ensemble search algorithm to the function $h$ results in a point that returns the value zero.  It is easy to see that, by construction, this point belongs to the basin of attraction of the global minimum.  We return then to the original function $f$ and apply the descent technique of choice that will lead to the global minimum.  If the basin of attraction of the global minimum is narrow, the gradients of the function $f$ may  reach very large values which may cause overshots.  Once that phase of the algorithm is reached, one can apply a scaling (dilation) transformation that maintains the descent mode but moderates the gradients.  On the other hand, as one approaches the global minimum, the gradients become very small and certain acceleration techniques based on non-Lipschitzian dynamics may be required \cite {BP, BPR}.  If the global minimum is attained at the boundary of the domain, the algorithm above will find it without additional complications.

\section{Br\"uschweiler's Ensemble Search Algorithm}
\label{Ensemble Algorithm}

Following Grover and Shor's breakthrough algorithms for quantum computing \cite{grover,shor,QC}, an alternative paradigm for computing has been suggested by Madi, Br\"uschweiler, and Ernst, which operates on \textit{ensembles} i.e., mixed states of identical spin sytems, using a spin Liouville space formalism for density operators \cite{Madi98}. The Liouville space for a system of $n$ spins contains the $n\times n$ density matrices representing all the possible quantum states of the spins. Ensemble algorithms are not quantum algorithms, insofar as they do not involve entanglement of quantum states, so they only use a small subset of the Liouville space. Throughout this paper, ``mixed states" describe a statistical ensemble of spins, rather than individual spins, so that each element of the ensemble performs part of the computation, in the same way as each processor in a classical parallel computer.

This new paradigm exploits the parallelism offered by simultaneously acting on many different input states in an ensemble of spins. In contrast, quantum computing with pure states relies on the parallelism of entangled states, to perform operations in a Hilbert space which is the tensor product of multiple qubits.

The advantage of ensemble algorithms is that they may be exponentially faster than the equivalent quantum algorithms, \textit{for adequate measurement sensitivities}. Two realizations of ensemble algorithms are provided by Br\"uschweiler's search algorithm \cite{Bruschweiler00} and the summing algorithm proposed by the authors \cite{Dhelon02}, which illustrate the trade-off between memory and speed capabilities.

Br\"uschweiler's algorithm for searching an unsorted database, uses the ensemble computing paradigm in the context of NMR technology. The strategy employs binary partition of the $N$ elements in the database, to find the desired element after $O(\log_2 N)$ oracle queries. We note that the lower bound derived by Nayak and Wu \cite{Nayak98} for the query complexity of quantum search algorithms, does not apply here, since the quantum lower bound  \cite{Beals98} used in their derivation assumes that the initial state is a pure state.

Ensemble computing requires an exponentially larger set of physical resources to encode the same number of distinct input states compared to quantum computing with pure states. On the other hand, ensemble computing holds the important advantage that it is insensitive to the decoherence time of the spins, which is an outstanding limiting factor for quantum computations involving entangled states.

A ``divide and conquer" scheme is used to test whether the special input $q$ belongs to exponentially finer and finer partitions of the set of input values. The algorithm is envisaged for a physically realizable system consisting of binary-valued spins, therefore the number of spins needed in the input register varies from $n$ to $1$, to accommodate the number of input values in partitions ranging in size from $N/2$ values down to $2$ values, respectively. It is also assumed that a 1-spin output register is available, to encode the value of $h \in \{0,1\}$.

The first step of the algorithm is to divide the input values into two equal partitions, $\chi_1^-=\{1,\dots,N/2\}$ and $\chi_1^+=\{N/2+1,\dots,N\}$, and test whether the special input $q$ belongs to $\chi_1^-$ or $\chi_1^+$. The numerical subscript, $k=1$, indicates the size of the partition as a fraction ($1/2^k$) of the entire set of input values, $\chi=\{1,\dots,N\}$, and the superscript $\pm$ differentiates between the lower and upper partitions.

The $n$-spin input register is initialized as an ensemble, namely an equally-weighted mixed state,
\begin{equation}
\rho_{in}^{(n)}=\frac{1}{N/2}\sum_{i \in \chi_1^-}|i\rangle_n\langle i|_n,
\label{rho_in}
\end{equation}
which accounts for half of the $N=2^{n+1}$ possible input configurations. The density operator $\rho_{in}^{(n)}$ can be represented in Liouville space by a density matrix that has non-zero elements only on its diagonal. The off-diagonal elements are all zero, indicating the absence of quantum coherence between any of the states $|i\rangle_n$. The ket states $|i\rangle_n$ correspond to the eigenstates of the Zeeman Hamiltonian created by a strong external magnetic field \cite{Ernst87}, and can also be written in terms of individual spins,
\begin{equation}
|i\rangle_n=|a_{i1}\rangle\otimes|a_{i2}\rangle\otimes\dots\otimes|a_{in}\rangle
\end{equation}
where $(a_{i1}, a_{i2},\dots,a_{in})\in \{0,1\}$ are the digits of the number $(i-1)$ in binary format, and $\otimes$ denotes the tensor product between the states.  The bra states, $\langle i|_n$, are the adjoint of the ket states. For further convenience, the state $|0\rangle$ denotes a spin ``up" and the state $|1\rangle$ denotes a spin ``down".

Each of the $N/2$ input configurations in the partition $\chi_1^-$ is realized by a sub-ensemble which contains many identical molecules with $n$ spins. The number of molecules in each sub-ensemble may vary, but the higher it is the more robust the algorithm is going to be against random errors. Ideally, all of the molecules in each sub-ensemble are initially set to one of the input configurations $|i\rangle_n\langle i|_n$, and the entire sub-ensemble responds to external controls as one ``mega-spin''.

At room temperature, the thermal equilibrium state of an $n$-spin ensemble in an NMR experiment is equal to the sum of $\rho_{in}^{(n)}$ and a traceless density matrix, with zero off-diagonal terms \cite{Gershenfeld97}. The diagonal elements of the traceless density matrix are given by permutations of Boltzmann factors, which have small magnitudes with respect to $2/N$, therefore the thermal equilibrium state closely approximates the desired initial mixed state $\rho_{in}^{(n)}$.

%We note again that the initial mixed state $\rho_{in}^{(n)}$ describes a statistical ensemble, thus the initialization is equivalent to assigning each of the input values $i=1,2,\dots,N/2$, to one of $N/2$ processors in a classical parallel computer.

The original approach proposed in Br\"uschweiler's search algorithm requires that the output register is initially set to zero. This requirement cannot be satisfied in an NMR implementation, as it is impossible to set all the spins in the output register to $|0\rangle$. However, Bruschweiler's search algorithm is still applicable, by taking the initial state to be the natural state available in NMR: a thermal mixture of $|0\rangle$ and $|1\rangle$ spins. Thus the entire ensemble (input and output registers) is described by $\rho_{in}^{(n)}\otimes\rho_{out}$,
\begin{equation}
\label{initial_state}
\rho_{in}^{(n)}\otimes\rho_{out}=\frac{1}{N/2}\sum_{i \in \chi_1^-}|i\rangle_n\langle i|_n\otimes\frac{1}{2}\{|0\rangle\langle 0|+|1\rangle\langle 1|\}.
\end{equation}
The effect of this modification is that measurements of the output register need to distinguish deviations from a thermal mixture, rather than from a uniform ensemble of $|0\rangle$ spins, following application of the search algorithm.

The function $h$, analogous to the oracle in Grover's search algorithm, is evaluated by applying a reversible unitary transformation $U_h$ defined as:
\begin{eqnarray}
U_h |i\rangle_n\otimes|0\rangle & \rightarrow & |i\rangle_n\otimes|h(i)\rangle \\
U_h |i\rangle_n\otimes|1\rangle & \rightarrow & |i\rangle_n\otimes|1\rangle .
\end{eqnarray}
The transformation has no effect on the eigenstates $|i\rangle_n$, but partitions the output register into two sub-ensembles: $1/2-1/N$ of the output spins have the value $|0\rangle$, and $1/2+1/N$ of the output spins have the value $|1\rangle$. The action of $U_h$ on the initial mixed state can be written as
\begin{equation}
\label{final_state}
U_h (\rho_{in}^{(n)}\otimes\rho_{out}) U_h^\dagger = \frac{1}{N/2}\sum_{i \in \chi_1^-}|i\rangle_n\langle i|_n\otimes\frac{1}{2}\{|h(i)\rangle\langle h(i)|+|1\rangle\langle 1|\}. \nonumber
\end{equation}

%represented by the density matrix:
%\[\left(
%\begin{array}{cccc}
%|0\rangle\langle 0| & 0 & \cdots & 0 \\
%0 & |0\rangle\langle 0| & & \vdots \\
%\vdots &  & \ddots & \\
%0 & \cdots & & |1\rangle\langle 1| \\
%\end{array}
%\right)\]

The transformation $U_h$ can be described \cite{Bruschweiler00} as a permutation of all states spanned by the input and output registers, which is applied to the system, to evaluate the function $h$ \textit{simultaneously} on all of the input values $i$ given by the initial state in Eq.(\ref{initial_state}). This is equivalent to evaluating the function $h$ concurrently on $N/2$ processors in a classical parallel computer. In a physical implementation, $U_h$ is the product of a sequence of fundamental unitary transformations for each of the spins in the output register.

The effect of the transformation $U_h$ is to create a small imbalance in the initial thermal state of the output register. If we can detect this imbalance reliably, then the special input value we are searching for is contained in the partition $\chi_1^-$.

In the physical implementation of the algorithm, the output register generates an output signal proportional to the number of spins in state $|1\rangle$. Hence the output signal will increase, if $\chi_1^-$ does contain $q$, or it will remain the same, if $\chi_1^-$ does not contain $q$. The latter case implies that $\chi_1^+$ \textit{must} contain $q$. This ensemble measurement is an analog process, which performs a single concurrent evaluation, and serves as a definite test of whether $q$ belongs to $\chi_1^-$ or $\chi_1^+$. Since only one out of $N/2$ input values \textit{may} change the output register, under the transformation $U_h$, the corresponding increase in the output signal is rather small (proportional to $2/N$).

Based upon this measurement, the partition containing $q$ is selected and then subdivided into two equal partitions, $\chi_2^-$ and $\chi_2^+$. The equally-weighted mixed state $\rho_{in}^{(n-1)}$ corresponding to the input values in $\chi_2^-$ is prepared in an $(n-1)$-spin input register,
\begin{equation}
\rho_{in}^{(n-1)}=\frac{1}{N/4}\sum_{i \in \chi_2^-}|i\rangle_n\langle i|_n,
\end{equation}
and the same approach outlined above is followed, to test if the special input $q$ belongs to $\chi_2^-$ or $\chi_2^+$. We note that, from an experimental point of view, it could be more convenient to prepare the same equally-weighted mixed state in the input register, regardless of which partition actually contained $q$, and then to use this information to determine an equivalent transformation that realizes the oracle for the appropriate partition.

Each test to determine the partition that contains $q$, reduces the size of the set of input values that contains $q$ by a factor of 2, therefore the special input value will be determined by a series of $n$ tests, i.e. the query complexity of the algorithm is $O(\log_2 N)$. The final test involves two equal partitions, $\chi_n^-$ and $\chi_n^+$, each containing a single input value, one of which is $q$.

For the first partition test, it is necessary that the measurement be able to distinguish changes in the output signal with a precision equal to or better than $1/N$. This requirement may be gradually relaxed for subsequent tests. Indeed, at the $k$-th test, the precision of the measurement only needs to be equal to or better than $1/2^k$, and by the final test, which determines whether $q$ belongs to $\chi_n^-$ or $\chi_n^+$, the precision has to be equal to or better than $1/2$. Of course, setting the precision to the minimum required level, $1/2^k$, for the $k$-th partition test, may provide some practical advantages in an experimental implementation, though it does not lower the query complexity.

An additional limitation on the measurement sensitivity for the output register is the inevitable noise present in measurements, especially due to thermal fluctuations in an NMR implementation. As the number of sample points $N=2^{n+1}$ increases, the difference between distinct normalized output signals in the first partition test will eventually become comparable to the 
noise level. In order to reliably measure the maximum normalized output signal $1/2^n$ in the first partition test, we require an error level $\Delta_1 \leq 1/2^{n+1}$. If the error level for a single experimental trial is $\Delta_1>1/2^k$ ($k\leq n$), we cannot reliably measure the signal from the output register for any of the first $k-1$ partition tests, because significant differences, which are smaller than $1/2^k$, cannot be detected reliably.

However the error can be reduced by repeating the algorithm a number of times. The error level scales inversely with the square-root of the number ($N_e$) of experimental trials, $\Delta_{N_e} \propto 1/\sqrt{N_e}$, because the variance in the measurement is inversely proportional to $N_e$. This implies that to successfully implement the first partition test, we have to repeat it at least $N_e \approx 2^{2(n+1-k)}=(N\Delta_1)^2$ times. Thus the single-run query complexity of the algorithm is $O(\log_2 N)$, but the overall query complexity becomes $O(N_e \log_2 N)$, due to the repeated trials. When $k\sim n$, $N_e\approx O(1)$, so the query complexity remains $O(\log_2 N)$. However, in the worst case, $k\ll n$, the query complexity increases to $O(N^2 \log_2 N)$.

\section{Discussion}
\label{Discussion}

%\subsubsection{The Optimization Problem}

\textit{The Optimization Problem.}  We have shown that a broad class of GOPs can be solved in a constant time, with a $guarantee$, by exploiting available additional information about the objective function. Our approach enables us to map the GOP into a discrete search, and \textit{guarantees} to find a solution for the global minimum in a constant time, by efficiently eliminating all basins of attraction except for the basin of attraction that contains the global minimum.

The additional information obtained by satisfying the assumptions (i)-(iii) in Section \ref{Mapping}, is readily available for some notoriously hard GOPs, such as the golf course problem. In general, a certain amount of information is usually available in every application, but present algorithms either do not or cannot use it. Sometimes, this information may be difficult to obtain, but its benefits significantly outweigh its cost. Indeed, for high-dimensional, computationally intensive problems, the advantage of reducing the complexity of the problem from exponential to polynomial is extremely significant. On the other hand, if no information is available at all, then a solution for the most general type of continuous GOP cannot be guaranteed in finite time.

In principle, the conditions (i)-(iii) imposed on the functions $f$ in Section \ref{Mapping} are satisfied can be relaxed to some extent.

Assumption (i) is satisfied by a large class of important practical problems, namely parameter identification, pattern recognition, and other classes of inverse problems.  In these problems, the absolute minimum, namely zero, is attained for the correct values of the parameters, matching of patterns, and fitting of output to input.  Assumption (i) can be relaxed in the sense that the function may have multiple global minima, all equal to zero.  Functions with multiple global minima will simply result in search problems with multiple ``special" elements and can be treated accordingly.
  
Assumption (ii) can be replaced with the much weaker and altogether reasonable assumption that $f$ has a \textit{finite} number of local minima. This would prevent the value of any local minimum to be infinitesimally close to the value of the global minimum.

Assumption (iii) is the most difficult to relax in practice, since this  information is crucial in ensuring the \textit{high efficiency} and \textit{guaranteed success} of the algorithm. In fact, without additional information about the size of the basin of attraction, there is no algorithm that - applied on a general problem - can \textit{guarantee} to find the global minimum in reasonable time. Indeed, if the discretization cell is much smaller than the size $\textbf{r}_\delta$ of the basin of attraction, many input values will return the output zero upon the application of the oracle $U_h$. In order to determine one of these input values with certainty, the ensemble search algorithm has to be applied for the same number of steps used to find a single special input value. Any of the determined values belong to the basin of attraction and will eventually lead to the global minimum. On the other hand, if the discretization cell is larger than $\textbf{r}_\delta$, there is a chance to miss the basin of attraction altogether. Moreover, if the basin of attraction is infinitesimally narrow, any algorithm would require an unbounded number of evaluations.\\

%\subsubsection{The Search Algorithm}

\textit{The Search Algorithm.}  The overall query complexity of the ensemble search algorithm used in this paper is always less than the overall query complexity for Grover's search algorithm implemented using NMR pseudopure states \cite{Gershenfeld97}, which is $O(N^2 \sqrt N)$. However, the ensemble search algorithm is less efficient than the (theoretical) implementation of Grover's search algorithm using pure states, for database sizes above a certain threshold value, which is determined by the error level. Br\"uschweiler \cite{Bruschweiler00} estimates that his algorithm is more efficient only for databases of size $N$ fulfilling the condition
\begin{equation}
N \sqrt N \log_2 N < \Delta_1^{-2}.
\label{N_max}
\end{equation}
We note that it is appropriate to make the above comparisons, since adding more quantum processors in an attempt to parallelize Grover's search method, does not achieve any further reduction in complexity.

In addition, ensemble algorithms do not require quantum coherence, thus they are more robust and demand less additional resources than quantum algorithms. For example, in an NMR implementation, quantum algorithms need to finish within the $T_2$ decoherence time of the system. This requirement is much more stringent than that for NMR ensemble algorithms, which only need to finish within the thermal relaxation time $T_1 \approx 50$-$1000$ $T_2$ \cite{Gershenfeld97,Ernst87}.

To date, Br\"uschweiler's search algorithm has been implemented experimentally by two separate groups \cite{Xiao02,Yang02}, both using a 3-qubit NMR ensemble system. While a database size up to $N=8$ is not very impressive, these experiments have demonstrated that the algorithm is based on sound principles. Both groups have improved on the original form of the algorithm, by either using measured spectra \cite{Xiao02} to read out the ensemble average of the output register, or by completely removing the need for the output register \cite{Yang02}, and using a modified oracle transformation, so that the algorithm becomes irreversible. These modifications make the algorithm more robust, and easier to realize, respectively, but they do not change the query complexity results we presented in Section \ref{Ensemble Algorithm}. Other searches in NMR ensemble computing have also been proposed and implemented \cite{Xiao02b,Khitrin02}.

However, we note that a number of important experimental issues remain to be addressed, before the algorithm can be scaled up to very large numbers of spins. These issues include finding suitable molecules with $n>3$ nuclear spins, designing the sequence of pulses that implements the oracle transformation $U_h$ efficiently, given that the interaction of remote nuclear spins is very weak, and enhancing the measurement sensitivity for large ensemble systems. The point of this proposal is proof-of-principle, which may eventually lead to something practical once NMR technology develops significantly.

The SNR for a single measurement is currently $10^4-10^7$ depending on the type of nuclei used \cite{Chuang98}, and the strength of the applied magnetic field. Taking the best corresponding value for the error level, $\Delta_1\approx 10^{-7}$, the ensemble search algorithm is more efficient than Grover's search algorithm using pure states, for database sizes up to $N \approx 10^8$, which corresponds to an input register of $n\approx 27$ qubits. This threshold size is already of practical interest, and will increase in the future, as the error level decreases with advances in NMR technology. The enhancement of measurement sensitivity in NMR implementations is presently being investigated by a large number of research groups, which are aiming to decrease the error level by some constant factor. In the future, high-sensitivity methods from other fields such as optical sensing of ultra-weak signals \cite{Liu02} using injection-locked lasers may also be adapted to NMR experiments.\\

%\subsubsection{Conclusion}

%Also, \textit{algorithmic cooling} \cite{Boykin02} via the polarization heat bath promises to enable large-scale NMR quantum computers, and it remains to be explored in the future whether a similar technique may be used to decrease the error level for ensemble computing.

%In future work, other conventional optimization methods e.g., simulated annealing and algorithm portfolios might also benefit from additional parallelism, and it will be of interest to compare our proposal with these methods.

\section{Conclusion}

In this letter, we have shown that additional information about the objective function can be used to reduce the complexity of the GOP to manageable levels, thus guaranteeing that the correct solution will be found. Conventional classical algorithms cannot take advantage of this information and quantum algorithms yield only a polynomial reduction in complexity. In contrast, ensemble algorithms hold the prospect of solving the continuous GOP in a constant time. For a restricted number of function samples, where the database size satisfies Eq.(\ref{N_max}), an ensemble search algorithm requires only $O(\log_2 N)$ invocations of the objective function $f$. However, this result is not asymptotic with respect to the input size of the GOP, due to the limitation imposed by Eq.(\ref{N_max}).

\section*{Acknowledgments}
This work was partially supported by the Engineering Research Program of the DOE Office of Basic Energy Sciences under contract DE-AC05-00OR22725 with UT-Battelle, LLC.

\end{document}